\documentclass[12pt]{article}
\usepackage{amsmath,amssymb}
\usepackage{graphicx}
\usepackage[english]{babel}
\usepackage{epsfig}

\usepackage[margin=1.2in]{geometry}
\usepackage{verbatim}
\usepackage{varwidth}
\usepackage{color}
\begin{document}

\title{Classical and relativistic $n$-body problem: \\ from Levi-Civita to the most advanced interplanetary missions}
\author{Sara Di Ruzza\thanks{This research has been supported by ERC project 677793 Stable and Chaotic Motions in the Planetary Problem. }}      
\date{\today}
\maketitle
\noindent Dipartimento di Matematica Tullio Levi-Civita \\
       Universit\`a degli Studi di Padova\\
       via Trieste, 63 - 35121 Padova, Italy      \\
            sdiruzza@math.unipd.it \\
              ORCID 0000-0003-4858-3535       

 \vskip1cm
 
\noindent {\bf Abstract}\\
The $n$-body problem is one of the most important issue in Celestial Mechanics. This article aims to retrace the historical and scientific events that led the Paduan mathematician, Tullio Levi-Civita, to deal with the problem first from a classic and then a relativistic point of view. We describe Levi-Civita's contributions to the theory of relativity focusing on his epistolary exchanges with Einstein, on the problem of secular acceleration and on the proof of Brillouin's cancellation principle. We also point out that the themes treated by Levi-Civita are very topical.  Specifically, we analyse how the mathematical formalism used nowadays to test General Relativity can be found in Levi-Civita's texts and evolves over the years up to the current Parametrized version of the Post-Newtonian approximation (PPN) which is used in high precision contexts such as important space missions designed also to test General Relativity and which aim to estimate with very high accuracy the PPN parameters.\\

\noindent {\bf keywords}\\
General Relativity, Celestial Mechanics, classical $n$-body problem,  relativistic $n$-body problem, PPN parameters, BepiColombo\\

\section{Introduction}
\label{intro}
\par\noindent
A wide part of the research of Levi-Civita, starting in Padua and then in Rome, was concerned with the classical and relativistic $n$-body problem. By this note, the author would like to mark the admirable line that takes together these two aspects, classical and relativistic, of this deep scientific matter. 
\par\noindent
Such researches, although are conceived, at the beginning, in the classical Newtonian context, are linked  as matter of fact also to the fundamentals of relativity. This topic, in recent years, appears to be extremely current and finds applications in space sciences. Just think to the BepiColombo space mission, an interplanetary mission joint by the European Space Agency (ESA) and the Japanese Aerospace Exploration Agency (JAXA) in order to explore Mercury. The mission has been launched on October 20, 2018 and the orbit insertion is expected in December 2025. One of the main goal of the mission is to offer further tests to General Relativity. A very realistic planetary model is used in order to write the Heliocentric dynamics of Mercury including gravitational perturbations as the $J_2$ of the Sun, the whole planetary and asteroid perturbations; moreover, relativistic effects are added in an approach that establishes the bases in the Levi-Civita theory. Then a sophisticated Orbit Determination is performed in order to determine the Post-Newtonian parameters with a very high accuracy which is well enhanced by subtle relativistic corrections (for details \cite{csdatm_orbit14,sddctm_BC}).

\smallskip

\par\noindent
The present historical-scientific paper is also a cultural consequence of the contribute of the author to the translation from French into Italian of Levi-Civita last article {\it \lq\lq Le probl\`eme des $n$ corps en relativit\'e g\'en\'erale\rq\rq} \cite{TLC1950} and it aims to connect different aspects of Celestial Mechanics. Moreover, the paper comes also from the fact that the Mathematics Department of the University of Padua has been recently dedicated to Tullio Levi-Civita and a carefully commented reprint \cite{TLC2018} of a selection of his articles has been presented in that occasion, together with a high quality reprint of the  original  handwritten thesis\footnote{recently rediscovered.} of Levi-Civita, elaborated under the supervision of his great master Gregorio Ricci-Curbastro.

\smallskip

\par\noindent
In Section~\ref{sec:2} we analyse some historical aspects of Levi-Civita's life that led him to approach the relativistic $n$-body problem, starting from his passion for geometry through the study of the classical $n$-body problem. In Section~\ref{sec:3} we show the main results and contributions of Levi-Civita in the relativistic $n$-body problem. Moreover we highlight the scientific relationships between Levi-Civita and his relativistic mathematical and physical colleagues. Finally, in Section~\ref{sec:4}, we present the relativity experiment in the BepiColombo space mission and the mathematical approach used in order to test General Relativity. 

\section{Some important historical events in the life of Tullio Levi-Civita}
\label{sec:2}

\subsection{From Differential Geometry to the $n$-body problem}
\label{sec:2.1}

\par\noindent
In the first few years of his university life, Tullio Levi-Civita approached Differential Geometry thanks to his mentor Gregorio Ricci-Curbastro. In 1894, Levi-Civita prepared, under Ricci-Curbastro supervision, his degree dissertation \lq\lq On Absolute Invariants\rq\rq, which represented the basis of a new branch of geometry that was called Absolute Differential Calculus. Then, the two mathematicians continued to be interested in Differential Geometry and in 1901, solicited by a prominent mathematical pioneer, Felix Klein, Ricci-Curbastro and Levi-Civita published the essay \lq\lq {\it M\'ethodes de calcul diff\'erentiel absolu et leurs applications}\rq\rq \ (Methods in absolute differential calculus and their applications) \cite{RCLC_methods} on one of the most important mathematical journals of that time, the Mathematische Annalen. The international scientific community did not react enthusiastically to this work: on the contrary, there was a great disinterest in the Absolute Differential Calculus. This involved that Levi-Civita was urged to look towards new horizons (for more details we refer to~\cite{TLC2019}). 
\par\noindent
Thus, in the first 15 years of the new century, Tullio Levi-Civita did not work on Absolute Differential Calculus, but his attention was mainly captured by Analytical Mechanics, Stability Theory and Celestial Mechanics and by problems connected with the newtonian $n$-body problem, in particular the three-body problem. This is the period in which he devised the \lq\lq three-body problem regularization\rq\rq \ ~\cite{TLC1904,TLC1906}, which is still the most efficient regularization available in the literature in the case of the planar restricted three-body problem.

\subsection{The approach to the relativistic $n$-body problem}
\label{sec:2.2}

\par\noindent
We wonder how Tullio Levi-Civita became interested in the relativistic $n$-body problem.
\par\noindent
In 1915, the classical physicist Max Abraham, a colleague of Levi-Civita, introduced him to the first version of the General Relativity of Albert Einstein and Marcel Grossman. Abraham's intent was that Levi-Civita demolished Einstein's theory. In fact, he asked to invalidate the theory of the brilliant german physicist. This request led Levi-Civita into a surprising and pleasant discovery: if most of the scientific community had accepted the Absolute Differential Calculus with disinterest, it had left an indelible mark on Albert Einstein. In fact, the theory of Absolute Differential Calculus, developed by Levi-Civita and his mentor Ricci-Curbastro, was used by Albert Einstein, as crucial mathematical scaffolding for General Relativity, his revolutionary theory of gravitation. In the first relativity of modern physics, envisioned by Galileo and Newton, the bodies' trajectories are each other deflected because of instantaneous forces generated by their masses, all such objects living in a Euclidean space. Conversely, in Einstein's relativity, the trajectories are nothing but geodesics --namely, paths of minimal local length-- in a non-Euclidean geometry of the space-time, a geometry which in turn is shaped by the masses. Levi-Civita read Einstein's work with great interest and was fascinated by his theory. He found some inaccuracies that concerned with local coordinate transformations. In 1915, a dense correspondence followed between Levi-Civita and Einstein, in which the german physicist strongly defended his equations from the meticulous observations pinpointed by Levi-Civita. The exchange of letters was interrupted in May 1915, when Italy entered in the World War I. A few days before, Einstein had acknowledged his error: in a letter he expressed his heartfelt gratitude to Levi-Civita, whom he considered a friend and the two men will be connected in a sincere friendship. 
\par\noindent
This led to two important consequences in the life of Levi-Civita: he rediscovered his love for geometry and he remained inexorably linked to the development of the Theory of Relativity.
\par\noindent
After exchanging letters with Levi-Civita, Einstein was able to improve and complete his theory and he proposed a revolutionary geometrization of gravitational physics: the field generated by the mass distribution starts to be interpreted as a curvature of space-time. In 1917, Levi-Civita sensed the universal reach of this approach and he was able to identify an important geometrical side: the concept of {\it Parallel Transport} in a curved space added to the notion of Riemannian geometry. As it turns out, the velocity of a geodesic evolves parallel to itself. Let us mention that Parallel Transport has proved quite fruitful, for it would eventually become the basic idea to Theory of Connections, a crucial branch of modern geometry. The Parallel Transport Theory represented one of the highest goals for Levi-Civita \cite{TLC1917}.
\par\noindent
In the following years, it was therefore a natural consequence that Levi-Civita dealt with the relativistic $n$-body problem.

\subsection{A disseminator of Relativistic Theory}
From the $1^{st}$ January 1919 Tullio Levi-Civita began to work at the University of Rome and he did not find colleagues particularly interested in relativistic theories. In March he gave a lecture entitled {\it \lq\lq Come potrebbe un conservatore giungere alla soglia della nuova meccanica\rq\rq}   (How could a conservative reach the threshold of new mechanics) \cite{TLC_1919}. The title and its initial eight lines will touch even the most inexperienced mathematician: Levi-Civita, the progressive mathematician, defends conservatism as a premise for progress, a task not to be underestimated by those who try to advance in research. He realised this by demonstrating that the evolution of classical mechanics into general relativity is a necessary process to explain certain experimental confutations, although the classical vision can be recovered as an approximation of the new theory for much smaller velocities than light velocity. Levi-Civita addressed to the dissemination of the Theory of Relativity in Italy and he became the mathematician who disseminates General Relativity, by writing articles and essays on it.

\section{Levi-Civita contribution to the relativistic $n$-body problem}
\label{sec:3}
\par\noindent
Although it has not been studied extensively from a historical point of view, it is known that from the beginning of the 1930s Levi-Civita remains deeply involved in the relativistic $n$-body problem. Two recent texts by Nastasi and Tazzioli \cite{nastasi_tazzioli2005} and by Goodstein \cite{Goodstein2018} provides important information on the genesis and evolution of the interest of Levi-Civita to this topic, while previously we have some information from the important commemoration in 1946 by his collaborator and friend Ugo Amaldi \cite{amaldi46} and subsequently by a long article by Andr\'e Lichnerowicz \cite{Lichnerowicz_1973}.
\par\noindent
In the 1920s, figures like Albert Einstein, Wilhelm de Sitter, Johannes von Droste and Karl Schwarzschild had already achieved important results on the relativistic problem of the two bodies (motion of an infinitesimally small mass in the field of a given spherical mass) before, and on the relativistic $n$-body problem then. In the Lichnerowicz article just mentioned, Marcel Brillouin explains that one of the main difficulties was integrating non-linear field equations, and solving the problem of singularities of self-interaction, inevitable in field theory. In fact, it is not possible to simplify the equations of motion by introducing a relativistic principle that plays the same role of the Newtonian action and reaction principle of Classical Mechanics. The need to develop a general approximation method emerged in order to produce explicit solutions within a prefixed order.
\subsection{The problem of the secular acceleration}
\par\noindent
In 1937 Levi-Civita published his crucial article {\it \lq\lq The relativistic problem of several bodies\rq\rq} \ in the American Journal of Mathematics \cite{TLC1937a}. It was basically the content of his lecture at the \lq\lq Harvard Tercentenary Conference of Arts and Sciences\rq\rq \ held in September 1936. It contained an observation by Einstein in the Appendix  concerning the energy tensor ${\textsf T_{ik}  }$ in Einstein equations 
\begin{equation*}
	{\textsf R_{ik} }-\frac{1}{2}R g_{ik}  = \frac{8 \pi G}{c^4} {\textsf T_{ik}} \,\,\,\,\, (i,k = 0,1,2,3) \ ,
\end{equation*}
where ${\textsf R_{ik} }$ is the Ricci curvature tensor, $R$ his trace, $c$ the light velocity and $G$ the Cavendish constant. The $g_{ik}$ are the coefficients of the element $ds^2$ of the space-time metric. Einstein expressed his doubts about the omission by Levi-Civita of a term that represents isotropic pressure inside the energy-momentum tensor $ {\textsf T_{ik}}$. This term will then be reintroduced by Levi-Civita in his latest work {\it \lq\lq Le probl\`eme des $n$ corps en relativit\'e g\'en\'erale\rq\rq} \cite{TLC1950} (posthumously published in 1950). Lichnerowicz explains to us how, in this article, Levi-Civita had coherently deduced the differential system that describes the motion of the centres of gravity of the $n$-bodies that takes into account the main parts of relativistic corrections. 
 \par\noindent 
 In the same year (1937) and in the same journal, Levi-Civita also published {\it \lq\lq Astronomical Consequences of the Relativistic Two-Body Problem\rq\rq} \cite{TLC1937b}, a text of one of his lectures held at Harvard University, where he applied his method and achieved remarkable results on the so-called secular acceleration. In this work dealing with the 2-body problem, Levi-Civita found a secular acceleration linked to the center of mass of the considered binary star system and to the periastron with respect to the star with greater mass, acceleration which resulted not to cancel.
  \par\noindent
  The problem of secular acceleration is precisely the cause of an important exchange of letters between Levi-Civita and Arthur Stanley Eddington, who was a well-known english astrophysicist who did extensive research on the Theory of Relativity. He played an important role in scientific dissemination, and in particular, thanks to him, Einstein's new theory came to England. One of his major achievements was his measurement of the deviation of light during an eclipse that represented an important goal of General Relativity.
\par\noindent
Eddington manifested perplexity about the existence of the secular acceleration found by Levi-Civita: his calculations were in fact different from those of Levi-Civita. For this reason, in a letter written in March 1938, Eddington invited Levi-Civita to explain all the passages. Here is an excerpt from the letter: \\
{\it \lq\lq Dear Professor Levi-Civita, \\
I am sending you a copy of a paper, which I am proposing to publish in the Proceedings of the Royal Society, which treats the problem of the secular acceleration. As you know I was very interested in the problem. I have an exceptionally able student G.L. Clark, and some of the most vital parts of the paper, especially the discovery of de Sitter's error, are due to him (...)\rq\rq}. 
 \par\noindent 
The paper he referred to is {\it \lq\lq The problem of n bodies in general relativity theory\rq\rq} \cite{ce_1938}, where we can read: \\
{\it \lq\lq In a recent investigation of the problem of two bodies in general relativity theory, Prof. Levi-Civita (1937b) has reached the conclusion that the centre of gravity has a secular acceleration in the direction of the major axis of the orbit towards the periastron of the larger mass\rq\rq}. 
  \par\noindent        
Levi-Civita's response was immediate, but he specified that he needed more time to review his calculations. Robertson and Clark were both determined to control Levi-Civita's calculations. Coincidentally, they met at Princeton and by joining their efforts they managed to find the error. They immediately saw that Levi-Civita had based the article \cite{TLC1937b} on De Sitter's errors (this fact is also cited in an article by Clifford Will \cite{clifford2011}). Levi-Civita recognised the error and was grateful to the two physicists. \\
The correct method of Levi-Civita was published only in 1950, when the volume of the {\it M\'emorial} \cite{TLC1950} was finally printed. 
\par
\begin{figure}
  \includegraphics[width=0.5\textwidth]{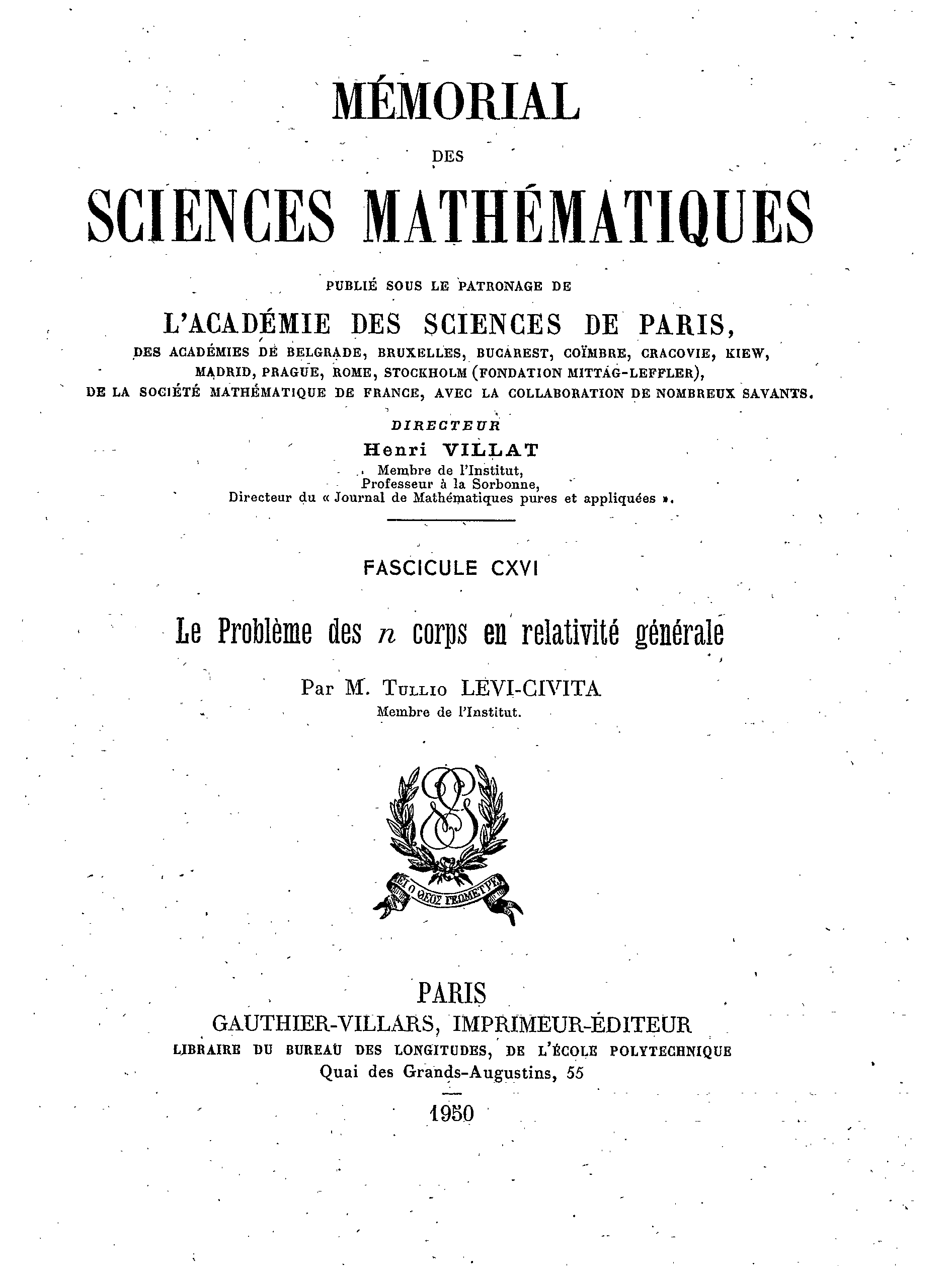}
  \includegraphics[width=0.5\textwidth]{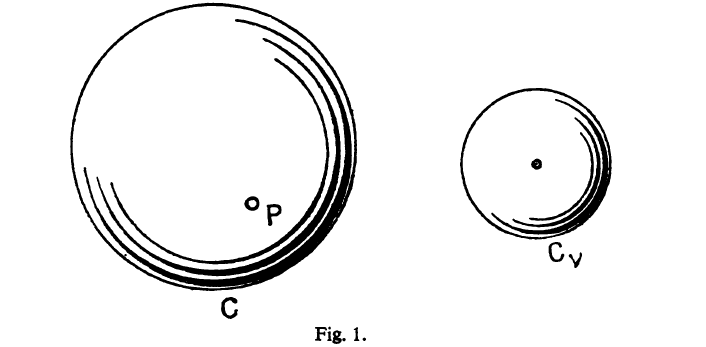}
\caption{On the left, the cover of {\it Le probl\`eme des $n$ corps en relativit\'e g\'en\'erale} \cite{TLC1950}; on the right, the only figure in this book representing the two gravity  interacting  bodies.}
\label{fig:1}      
\end{figure}

\noindent       
In the Preface of the aforesaid volume, Levi-Civita, referring to the secular acceleration, reiterates that {\it \lq\lq This hasty conclusion came only from a material error of calculation (the passages presented in this volume have nevertheless been perfectly corrected)\rq\rq}. \\In Fig.~\ref{fig:1}, the cover and a figure of \cite{TLC1950} are shown.
\subsection{The demonstration of the \lq\lq cancellation principle\rq\rq}  
In any case, as Lichnerowicz observed, Levi-Civita was right to think that his method was worthy of publication: it was really more fruitful and natural than Robertson's method, as shown also by Vladimir Fock in his works published in 1939 and 1941 \cite{fock1939,fock1941}, containing the final version of the new relativistic mechanics. Moreover, in the last work of Levi-Civita, the definitive demonstration of the so-called \lq\lq cancellation principle\rq\rq \ that Marcel Brillouin mentioned before emerges: each body in the system exerted no influence on the motion of its centre of gravity so that the gravitational self-interaction of the bodies does not affect the chosen approximation order.
 \par\noindent      
To deal with the $n$-body problem, Levi-Civita used the so-called PN (Post-Newtoni\-an) approximation, based on the assumption that the gravitational fields are sufficiently weak and that the characteristic motions of matter are slow, compared to the light velocity. This approximation is non-linear  and is therefore more sophisticated than the linearised versions of the theory.
 \par\noindent  
The theoretical model built with the Post-Newtonian method is confirmed in the experimental results and in the astronomical measurements and is still used today. The inexplicable effectiveness of the Post-Newtonian theory, as the physicist Clifford M. Will rightly predicted (we refer again to \cite{clifford2011}), would have played an important role even in the identification of gravitational waves. Moreover, important space missions, such as Cassini in the past or BepiColombo in the future, are intended to test (thus confirm or deny) General Relativity, in its Post-Newtonian formulation. 

\section{Relativistic approach and results of BepiColombo mission}
\label{sec:4}
 \par\noindent  
 Let us show how BepiColombo space mission represents an extraordinary application of the relativistic $n$-body problem. One of the most important goals of the mission is in fundamental physics and in particular is to test General Relativity in its Post-Newtonian formulation. The possibility of carrying out a relativity experiment on the BepiColombo mission to Mercury was born from an idea by Andrea Milani and Luciano Iess from the early stages of the mission \cite{milani2002,milani2010,iess2001}.
 \subsection{The Parametrized Post-Newtonian formalism}
   \par\noindent  
 Equations of motion for celestial bodies used for Orbit Determination has to be compliant with General Relativity by considering the properties of light propagation in a curved space-time. An important contribution in the study of evolution of a system of heavy point particles interacting gravitationally through equations satisfying general relativity equations can be found in \cite{zeldovic65}.  It has been verified (see \cite{dicke1960}) that gravitation is a phenomenon of curved space-time and it must be described by a \lq\lq metric theory\rq\rq \ of gravity. General Relativity is an example of metric theories of gravity. As we can read in \cite{will1993,will2014}, the postulates of metric theories assert that matter and non-gravitational fields do not interfere with one another. The only gravitational field that rules the equations of motion is the metric itself. The other fields that could be present in a theory can only help to generate the space-time curvature associated with the metric. Matter can create these fields and these together with matter can generate the metric, but they cannot directly interact with matter. Matter responds only to the metric.
\par\noindent
For this reason, a metric theory differs from another only by defining the metric and the equations of motion, namely how matter and possibly other gravitational fields generate the metric. The comparison between the different metric theories of gravity through experiments becomes simpler if we take slow motion, weak field limit. This approximation, known as the Post-Newtonian limit, is sufficiently accurate to include all the tests that can be performed in the near future in the Solar System.
\par\noindent
 The metric of a theory may differ from that of another theory by modifying the coefficients that multiply each term in the metric. By substituting each coefficient with an arbitrary parameter we get a \lq\lq super metric theory of gravity\rq\rq \ whose special cases (particular values of the parameters) are the Post-Newtonian metrics of particular theories of gravity. This super metric is called the {\it Parametrized Post-Newtonian (PPN) metric}, and the parameters are called {\it PPN parameters}. This use of parameters to describe the Post-Newtonian limit of metric theories of gravity is called the {\it Parametrized Post-Newtonian (PPN) formalism}. A primitive version of such a formalism was devised and studied by Eddington in \cite{eddington1922}, Robertson in \cite{robertson1962} and Schiff in \cite{schiff1967}; an important contribute was given by Nordtvedt in \cite{Nordtvedt_1968}. The Eddington-Robertson-Schiff formalism treated the Solar System metric as that of a spherical non-rotating Sun, and idealised the planets as test bodies moving on geodesies of this metric \cite{will1993}. A unified version of the PPN formalism was presented by Will and Nordtvedt in \cite{wn1972}. \\
 \noindent A very pioneering and influential contributions on relativistic celestial mechanics and N-body relativistic system dynamics is given by Brumberg (see \cite{brumberg91} based on his earlier influential book, Relativistic Celestial Mechanics \cite{brumberg72} in Russian), where the author describes the results of the general relativistic theory of motion of celestial bodies, both in case of natural bodies (as example, planets of the Solar System) and artificial bodies (Earth's artificial satellites). The author, also, deal with the relativistic theory of astronomical reference frames, time scales, and the reduction of observations: the information given by astronomical observations characterizes not only the object of observation but the observer as well. In order to use information obtained by different observers or even by one and the same observer but at different moments of time it is necessary to perform a reduction of the observations. Dealing with space missions and ephemeris astronomy, it is mandatory to develop a self-consistent theory of timescales based on relativistic theory of astronomical reference systems (for a well-developed theory see, for example, \cite{ashby79}.)

\subsection{The PPN approach applied to BepiColombo}
\par\noindent
  The equations of motions in the Orbit Determination of BepiColombo mission follow the Parametrized Post-Newtonian approach, in particular, in its Lagrangian formulation as it has been deducted by Moyer \cite{moyer1971,moyer2000}. The Post-Newtonian propagation of light rays may also be obtained using the above approximations to the metric. Since light moves along null trajectories, the Lagrangian $L$ must be formally identical to zero. In the first order Newtonian limit, this implies that light must move on straight lines.
\par\noindent
It is also token into account that the cancellation principle of Brillouin holds at the first Post-Newtonian approximation (as Levi-Civita proved in \cite{TLC1950}) in the sense that all large direct self-action effects cancel in the equations of motion, so that the final equations of motion can be written in terms of only centres of mass (see also \cite{battista_etal2017}). 
\par\noindent
The $n$-body point-mass relativistic equations of motion can be derived from the $n$-body point-mass metric tensor ${\textsf g_{ij}}$. The trajectory of a massless particle or a celestial body in the gravitational field of $n$ other celestial bodies is a geodesic curve which extremises the integral of the interval $ds={\textsf g_{ij}} dx^idx^j$, being the $x^i$ the space-time coordinates, between two points:
\begin{equation*}
	\delta \int ds= 0 \ .
\end{equation*}
In order to obtain the equations of motion in Solar System barycentric reference frame with time coordinate $t$ as independent variable, the previous equation is written as
\begin{equation*}
	\delta \int L dt = 0 \ ,
\end{equation*}
where the Lagrangian $L$ is obtained by detailing the term $\frac{ds}{dt}$.
\par\noindent
We would like to underline that the metric tensor used in \cite{moyer2000} is exactly the same used by in \cite{TLC1950}. In \cite{moyer2000}, the invariant interval $ds$ between two events with differences in their space and time coordinates is given in Equation (2--15) where the $n$-body metric tensor is given from Equations (2--1) to (2--6). Moyer introduces the two PPN parameters $\gamma$ and $\beta$. By substituting $\gamma =\beta =1$, as General Relativity theory proposes, Equation (2--15) in \cite{moyer2000} is exactly the same of Equation (III.63) in \cite{TLC1950} where the $n$-body metric tensor is given in Equation (III.60). Let us note that, by comparing the two different versions by Moyer and Levi-Civita, we have to take into account that Levi-Civita used completely different meaning in the definition of $\beta$ and $\gamma$ and, moreover, he does not multiply the time coordinate in the space-time reference system by the light velocity constant $c$. In the next section, we report the equations we are referring to.
\subsection{From Levi-Civita to Moyer equations}
\label{sec:4.3}
\par\noindent We want to show the equations given in \cite{moyer2000} and in \cite{TLC1950} for $n$-body metric tensor and the element $ds$. They are reported in the same form and with the same notations used by the two authors.\\
Let us start by Moyer formulation \cite{moyer2000}. The components of the $n$-body metric tensor ${\textsf  g_{pq}}$ are given from equations (2--1) to (2--6). Subscripts 1,2,3 refer to position coordinates, and 4 refers to time coordinate $t$ multiplied by the light velocity $c$; $\beta$ and $\gamma$ are the PPN parameters as in \cite{wn1972}. The equations for the components ${\textsf g_{pq}}$ are:
\begin{eqnarray}
\label{metric_tensor_moyer}
	{\textsf g_{11}} & = & {\textsf g_{22}}={\textsf g_{33}}=-\Bigg( 1 + \frac{2 \gamma}{c^2}\sum_{j \neq i } \frac{\mu_j}{r_{ij}} \Bigg) \nonumber \\
	{\textsf g_{pq} }& = & 0 \quad (p,q = 1,2,3; p \neq q) \nonumber \\
	{\textsf g_{14} }& = & {\textsf g_{41}} = \frac{2 + 2 \gamma}{c^3} \sum_{j \neq i } \frac{\mu_j \dot{x}_j}{r_{ij}}  \nonumber \\
	{\textsf g_{24}} & = & {\textsf g_{42}} = \frac{2 + 2 \gamma}{c^3} \sum_{j \neq i } \frac{\mu_j \dot{y}_j}{r_{ij}}  \\
	{\textsf g_{34}} & = & {\textsf g_{43}} = \frac{2 + 2 \gamma}{c^3} \sum_{j \neq i } \frac{\mu_j \dot{z}_j}{r_{ij}}  \nonumber \\
	{\textsf g_{44} }& = & 1 -\frac{2}{c^2} \sum_{j \neq i }\frac{\mu_j}{r_{ij}} + \frac{2 \beta}{c^4} \Bigg[ \sum_{j \neq i }\frac{\mu_j}{r_{ij}} \Bigg]^2 - \frac{1+ 2 \gamma}{c^4} \sum_{j \neq i }\frac{\mu_j \dot s_j^2}{r_{ij}} \nonumber \\
	&& + \frac{2(2 \beta -1)}{c^4}  \sum_{j \neq i }\frac{\mu_j}{r_{ij}}  \sum_{k \neq j }\frac{\mu_k}{r_{jk}} - \frac{1}{c^4} \sum_{j \neq i }\mu_j \frac{\partial^2 r_{ij}}{\partial t^2}  \nonumber
\end{eqnarray}
where the indices $j$ and $k$ refer to the $n$ bodies and $k$ includes body $i$, whose motion is desired. Moreover, $\mu_j$ is the gravitational constant for the body $j$, $r_{ij}$ is the mutual distance between body $i$ and body $j$, $\dot x_j, \dot y_j, \dot z_j$ are the velocity components of the body $j$ and $\dot s_j^2$ is the square of the norm of the velocity vector of the body $j$. \\
Setting $x^1 = x_i, x^2 = y_i, x^3 = z_i, x^4 = ct$ and substituting the components of the $n$-body metric tensor \eqref{metric_tensor_moyer} in the invariant interval $ds = {\textsf g_{pq}} dx^p dx^q$ between two events with difference in their space and time coordinates $dx^1, dx^2, dx^3, dx^4$, we get equation (2--15) of \cite{moyer2000}, namely:
\begin{eqnarray}
\label{ds_moyer}
	ds^2 & = & {\textsf g_{44}}\,  c^2 dt^2 + {\textsf g_{11}}(dx_i^2+dy_i^2+dz_i^2) \nonumber \\
	&& + 2\,  {\textsf g_{14} } dx_i c\,  dt + 2\, {\textsf g_{24}} dy_i c\,  dt + 2\,  {\textsf g_{34} }dz_i c \, dt \ .
\end{eqnarray}
 \noindent Let us continue by writing the $n$-body metric tensor appearing in \cite{TLC1950}. The notations are completely different from the previous ones. The component $x^0$ represents the time coordinate, while $x^1,x^2,x^3$ stand for the spatial coordinates. The terms $\beta$ and $\gamma$ have a different meaning which is written in the following formulas (no reference to the PPN parameters). The components of the $n$-body metric tensor given in equation (III:60) are:
\begin{equation}
\label{metric_tensor_TLC}
\left \{
\begin{aligned}
{\textsf g_{ik}} &= 0 \quad &&\bigl(i\neq k,i,k=1,2,3\bigr),\\
{\textsf g_{ii} }&=-\bigl(1+2\gamma\bigr) \quad &&\bigl(i>0\bigr),\\
{\textsf g_{0i} }&=4\gamma_i  \quad &&\bigl(i>0\bigr),\\
{\textsf g_{00}} &=1-2\gamma+2\gamma^2-2\zeta;
\end{aligned} \right.
\end{equation}
where
\begin{eqnarray}
\label{parameter_TLC}
\zeta & = & \varphi+\psi+\emph{v} \ , \nonumber \\
\gamma & = & \frac{f}{c^2}\int_S\frac{\mu\,dS}{r} \ ,  \nonumber\\
\gamma_i & = & \frac{f}{c^2}\int_S\frac{\mu\beta_i}{r}\,dS \quad \bigl(i=1,2,3\bigr) \ , \nonumber \\
\varphi & = & -\frac{f}{c^2}\int_S\frac{\mu\gamma}{r}\,dS  \ , \\
\psi & = & \frac{3}{2}\frac{f}{c^2}\int_S\frac{\mu\beta^2}{r}\,dS  \ , \nonumber \\
\emph{v} & = & \frac{1}{2}\frac{f}{c^2}\frac{\partial^2}{\partial x^{0^2}}\int_S\mu r\,dS \ , \nonumber
\end{eqnarray}
and where $S$ is the volume occupied by the $n$ bodies, $\mu$ is the mass density occupying a certain volume at a given time, {\it i.e.} $\mu= \eta/c^2$, where $\eta$ is the energy density according to the Einstain's concept of proportionality between mass and energy; $f \sim 6.675 \cdot 10^{-8} {\rm\,  g^{-1} \, cm^3 \, sec^{-2} }$ is the Gauss universal gravity constant and, finally, $r$ is the distance between two bodies.
Moreover,
 $$
 \beta_i=\frac{dx^i}{dx^0}\quad \text{and}\quad \beta^2=\sum_{i=0}^3\beta_i^2 \ .
 $$ 
 Thus, substituting the components of the metric tensor \eqref{metric_tensor_TLC} in $ds^2= {\textsf g_{ik}}dx^idx^k$, we get equation (III.63), namely:
 \begin{equation}
 \label{ds_TLC}
ds^2=\bigl(1-2\gamma+2\gamma^2-2\zeta\bigr)dx^{0^2}-\bigl(1+2\gamma\bigr)dl_0^2+8dx^0\sum_{i=0}^3\gamma_idx^i \ ,
\end{equation}
where
\begin{equation*}
dl_0^2=\sum_{i=0}^3dx^{i^2} \ .
\end{equation*}
With suitable substitutions and calculations Eq. \eqref{ds_moyer} turns out to be identical to Eq. \eqref{ds_TLC}.

\subsection{The relativistic experiment in BepiColombo mission}
\par\noindent
In the following we will refer to the notation and the equations derived in \cite{moyer2000}. Let us show how the PPN parameters can be introduced and which are their physical meaning. The amount of space-curvature which a standard mass produces in a given theory is measured by the parameter $\gamma$ while the parameter $\beta$ measures the non-linearity of Post-Newtonian gravity; they are called {\it Eddington} Post-Newtonian parameters and they are set equal to 1 in General Relativity. The parameter $\eta$, which plays a role in the definition of the inertial mass, is the {\it Nordtvedt} parameter, while $\alpha_1$ and $\alpha_2$ are the {\it preferred frame} parameters and in General Relativity holds $\eta = \alpha_1=\alpha_2 =0$; the meaning of these parameters will be clarified soon. In the equations of motion, two more parameters that are not precisely relativistic are introduced because they are strictly linked to the relativistic parameters; they are $J_{2 \odot}$, the solar quadrupole moment representing the currently constant dynamic oblateness of the Sun, and the gravitational parameter time variation $\zeta=\frac {d \mu_{\odot}}{dt}\frac{1}{\mu_{\odot}}$.
\par\noindent  
Let us define as $\mathbf x=(\mathbf{x_0}, \dots, \mathbf{x_N}) \in \mathbb R^{3(N+1)} $ and $\mathbf v=(\mathbf{v_0}, \dots, \mathbf{v_N}) \in \mathbb R^{3(N+1)} $, respectively, the barycentric positions and velocities of the $N+1$ bodies. The relativistic equation of motion is linearized with respect to the small parameters $v_i^2/c^2$ and $\mu_i/r_{ik}$ where $v_i$ is the norm of the barycentric velocity for each of the bodies of mass $m_i$, $c$ is the light velocity, $\mu_i= G m_i$ is gravitational parameter of the $i^{th}$ body and $r_{ik} = \| \mathbf{x_i}- \mathbf{x_j} \|$ is the mutual distance, appearing in the metric of the curved space-time.
In this framework, the relativistic Lagrangian of the planetary problem consisting of the Sun plus $N$ bodies is:
    \begin{equation*}
     L(\mathbf x,\mathbf v)= L_{NEW}+L_{GR0}+ (\beta-1) L_{\beta}+(\gamma-1) L_{\gamma}  +J_{2 \odot} L_{J_{2 \odot}}+L_{\alpha}+\zeta L_{\zeta} \ .
\end{equation*}
Let us describe in detail the explicit form of each term. The newtonian part is:
\begin{equation*}
        L_{NEW}= \frac{1}{2} \sum_{i=0}^N \mu_i {v_i}^2 +  \frac{1}{2} \sum_{i=0}^N \sum_{j \ne i} \frac{\mu_i \mu_j}{r_{ij}} \ ;
\end{equation*}
the portion without free parameters (apart $G$) at a zero order approximation is: 
\begin{eqnarray*}
    L_{GR0} & = & \frac{1}{8 c^2} \sum_{i=0}^N \mu_i v_i^4 +  \frac{1}{2c^2} \sum_{i=0}^N \sum_{j \ne i}\sum_{k \ne i} \frac{\mu_i \mu_j \mu_k}{r_{ij}r_{ik}} \\
    &&   + \frac{1}{2c^2} \sum_{i=0}^N \sum_{j \ne i} \frac{\mu_i \mu_j}{r_{ij}} \Big[\frac{1}{2}(v_i^2+v_j^2)- \frac{3}{2}(\mathbf{v}_i\cdot \mathbf{v}_j) \\
     && -\frac{1}{2}(\mathbf{n}_{ij}\cdot \mathbf{v}_i)(\mathbf{n}_{ij}\cdot \mathbf{v}_j)\Big] \  ,
\end{eqnarray*} 
where $ \mathbf{n}_{ij}= \frac{\mathbf{x}_i- \mathbf{x}_{j}}{r_{ij}}  $. The velocity-dependent modification of the two body interaction part  is ($\gamma-1) L_{\gamma} $, where
\begin{equation*}
    L_{\gamma}   =   \frac{1}{2c^2} \sum_{i=0}^N \sum_{j \ne i} \frac{\mu_i \mu_j}{r_{ij}} (\mathbf{v}_i- \mathbf{v}_j)^2 \ ;
\end{equation*}
the modification of the non-linear three-body general relativistic interaction part is ($\beta-1) L_{\beta} $, where
\begin{equation*}
      L_{\beta}    =   -  \frac{1}{c^2} \sum_{i=0}^N \sum_{j \ne i}\sum_{k \ne i}  \frac{\mu_i \mu_j \mu_k}{r_{ij}r_{ik}} \ ;
\end{equation*}
the contribution of the oblateness of the Sun is $J_ {2\odot} L_{J_{2\odot}}$, where
\begin{equation*}
   L_{J_{2\odot}}=  -  \frac{1}{c^2} \sum_{i=1}^N \frac{\mu_{\odot} \mu_i}{r_{0i}r_{ik}} \Big(\frac{R_{\odot}}{r_{0i}} \Big)^2 [3(\mathbf{n}_{0i} \cdot \mathbf{e}_0)^2 -1] \ ; 
\end{equation*}
with $R_{\odot}$ the Sun's radius and $\mathbf{e}_0$ the unit vector along the Sun's rotation axis. This term is not a relativistic one, but the high correlation between $J_{2\odot}$ and $\beta$ makes the study of this term necessary in this context.\\
Moreover, one goal is to measure also the time variation of the gravitational constant $G$, in fact, it is not possible to discriminate the change with time of $G$ from the change in time of $m_{\odot}$, thus a new parameter $\zeta=\frac {d \mu_{\odot}}{dt}\frac{1}{\mu_{\odot}}$ is introduced; the corresponding term in the Lagrangian is $\zeta L_{\zeta}$, where
\begin{equation*}
  L_{\zeta}=(t-t_0) \sum_{i \ne 0} \frac{\mu_{\odot} \mu_i}{r_{0i}} \ . 
\end{equation*}
Finally, the last term used in the Post-Newtonian formulation is that depending on the preferred frame parameters $\alpha_1, \alpha_2$, which can be written as  $  L_{\alpha}$:
\begin{equation*}
 L_{\alpha}=\frac{\alpha_2-\alpha_1}{4 c^2} \sum_j \sum_{i \ne j}\frac{\mu_i \mu_j}{r_{ij}} (\mathbf{z}_i\cdot \mathbf{z}_j) - \frac{\alpha_2}{4 c^2} \sum_j \sum_{i \ne j}\frac{\mu_i \mu_j}{r_{ij}}[(\mathbf{n}_{ij}\cdot \mathbf{z}_i)(\mathbf{n}_{ij}\cdot \mathbf{z}_j)] \  , 
\end{equation*} 
with $\mathbf{z}_i=\mathbf{w}+\mathbf{v}_i$ where $\mathbf{w}$ is the velocity of the Solar System barycentre with respect to the preferred frame, usually assumed to be the one of the cosmic microwave background, thus $|\mathbf{w}|= 370 \pm 10 \ km/s$ in the direction $(\alpha,\delta)=(168^{\circ},7^{\circ})$.
\par\noindent
The parameter $\eta$ is introduced in the model by substituting the gravitational mass $m_i^G$ with the inertial mass $m_i^I$
$$m_i^G=m_i^I(1+\eta  \Omega_0) \ $$
 where $\Omega_0=-3.52 \cdot 10^{-6}$ is the gravitational self energy of the Sun (recall that for tests of Strong Equivalence Principle (SEP) among the Solar System bodies, it is sufficient considering just the gravitational self energy of the Sun (see \cite{milani2002} for more details).
  For $\eta=0$ we have the SEP, where we do not discriminate between gravitational and inertial mass. If $\eta \ne 0$, then, we have a violation of the SEP.
\par\noindent
In BepiColombo mission, by a global least-squares fit, an accurate Orbit Determination (as treated in \cite{milanigronchi2010}) is performed and it allows to estimate the PPN parameters with very high accuracy. 
One of the first estimate of the parameter $\gamma$ and $\beta$ is given in \cite{schiff1967} where we can read that astronomical observation showed that $(1+\gamma)$ is equal to 2 with an uncertainty of roughly $20\%$ while the quantity $2(1+ \gamma) - \beta$ is equal to 3 with an uncertainty of about $1\%$. The actual estimates for $\gamma$ and $\beta$ are about an uncertainty of $2 \times 10^{-5}$ and $10^{-4}$, respectively \cite{ber_iess_tort_2003,will2014}. Last estimates presented in recent papers as \cite{dtms2016_nordtvedt,scdt_relativity,sddctm_BC,sstm2018,sdst2018}  prove that BepiColombo could give an accuracy up to $10^{-6}$ both for $\gamma$ and $\beta$. Until now, these last results have been obtained with simulated observations. Finally, BepiColombo is on its way to Mercury and in the coming years we will have real observations which allow to have new and very accurate estimates that will confirm or confute the values of the PPN parameters in General Relativity and provide new information on metric theories of gravity.

\section*{Acknowledgements}
I really thank Franco Cardin for introducing and involved myself in the interesting cultural journey of the Italian translation of Tullio Levi-Civita latest work \cite{TLC2019}. I would like to express my thanks also to Leonardo Don\`a with whom we have realised that work. Moreover, I would like to thank Francesco Fass\`o and Bruno Chiarellotto for their advices. Finally, I am very grateful to Gabriella Pinzari for her precious suggestions.

\section*{Compliance with Ethical Standards}

Conflict of Interest: The author declares that she has no conflict of interest.

%

\end{document}